\documentclass[twocolumn,english,preprintnumbers,amsmath,amssymb,pre,longbibliography]{revtex4-1}

\usepackage{graphicx}
\usepackage{color}

\begin{document}

\title{Nuclear quantum effects in structural and elastic properties of cubic 
       silicon carbide}

\author{Carlos P. Herrero$^1$\footnote{Electronic mail: ch@icmm.csic.es},
 Rafael Ram\'irez$^1$, and Gabriela Herrero-Saboya$^2$}

\address{$^1$Instituto de Ciencia de Materiales de Madrid,
           Consejo Superior de Investigaciones Cient\'ificas (CSIC),
           Campus de Cantoblanco, 28049 Madrid, Spain  \\
         $^2$CNR-IOM Democritos National Simulation Center,
           Istituto Officina dei Materiali, c/o SISSA,
           via Bonomea 265, IT-34136 Trieste, Italy}	 

\date{\today}

\begin{abstract}
Silicon carbide, a semiconducting material, has gained
importance in the fields of ceramics, electronics, and renewable
energy due to its remarkable hardness and resistance.
In this study, we delve into the impact of nuclear quantum
motion, or vibrational mode quantization, on the structural and
elastic properties of $3C$-SiC.
This aspect, elusive in conventional {\it ab-initio} calculations, 
is explored through path-integral molecular dynamics (PIMD)
simulations using an efficient tight-binding (TB) Hamiltonian.
This investigation spans a wide range of temperatures and
pressures, including tensile stress, adeptly addressing the
quantization and anharmonicity inherent in solid-state
vibrational modes.
The accuracy of the TB model has been checked by comparison with
density-functional-theory calculations at zero temperature.
The magnitude of quantum effects is assessed by comparing PIMD outcomes
with results obtained from classical molecular dynamics simulations.
Our investigation uncovers notable reductions of 5\%, 10\%, and 4\% in
the elastic constants
$C_{11}$, $C_{12}$, and $C_{44}$, respectively, attributed to atomic
zero-point oscillations. Consequently, the bulk modulus and Poisson's ratio
of $3C$-SiC exhibit reduced values by 7\% and 5\% at low temperature.
The persistence of these quantum effects in the material's structural 
and elastic attributes beyond room temperature underscores the necessity 
of incorporating nuclear quantum motion for an accurate description of 
these fundamental properties of SiC.
\end{abstract}

\maketitle

\section{Introduction}

Bulk silicon carbide is recognized as a semiconducting material with 
outstanding physical properties, such as low thermal expansion, high 
strength, thermal conductivity, and refractive index \cite{me07c}. 
It exists in more than 250 different polytypes, 
many of which have hexagonal or rhombohedral crystalline structures.
Moreover, there has been an increasing interest in other materials 
consisting of carbon and silicon, such as nanotubes, fullerenes, and 
two-dimensional structures \cite{me07c,hs11,sh15,ch21,sc-po23}.

Mechanical properties of bulk SiC have been intensively studied 
over the years using experimental techniques 
\cite{sc-zh13,sc-le82,sc-fe68,sc-la91} and theoretical approaches 
\cite{sc-va15,sc-ra21,sc-pe22,sc-sh00,sc-le15b,sc-wa96,sc-he23}, 
due to their significance in both basic research and various technological 
applications, including heat shielding, nuclear fuel particles, filament 
pyrometry, telescope mirrors, electric systems, and electronic devices.
In general, the behavior of semiconducting solids under high pressure 
conditions has recently received renewed attention as potential 
constituents of carbon-rich exoplanets, as extreme planetary interior 
conditions can significantly influence their physical properties 
\cite{sc-ni17,sc-ki22}.

One of the most prevalent forms of silicon carbide, stable under ambient 
conditions, is cubic $3C$-SiC. Extensive research has been conducted on 
this polytype, employing various computational techniques, with a notable 
focus on density-functional theory (DFT) calculations at $T = 0$ 
\cite{sc-le15b,sc-ki17,sc-sh18,sc-ra21}. 
Since accounting for finite temperatures requires the estimation of 
vibrational modes within a quasi-harmonic approximation 
\cite{mo05,de96,de19b,sc-pe22}, these methods are reliable at low
temperatures where anharmonic effects are relatively small, but
their accuracy can diminish due to increasing anharmonicity as the 
temperature rises.

Numerous studies investigating the finite-temperature characteristics 
of SiC have relied on classical atomistic simulations 
\cite{sc-sh00,sc-ka05,sc-iv07, sc-ku17,sc-ch21d,sc-ka21}. 
However, it is noteworthy that the Debye temperature $\Theta_D$ of 
silicon carbide greatly exceeds room temperature 
($\Theta_D \approx$ 1100~K for cubic SiC \cite{sc-zy96}). 
As a result, the interplay of nuclear quantum effects (or phonon quantization) 
and anharmonicities in the interatomic potential is expected to impact 
the material's physical properties at temperatures relatively high,
potentially on the order of, or even surpassing, room temperature.
The constraints posed by classical atomistic simulations 
can be surmounted by methods explicitly accounting for nuclear quantum 
motion. Such is the case for computational techniques grounded in 
Feynman path integrals \cite{gi88,ce95,br15,he16}. These techniques have 
gained prominence in recent years for exploring diverse material properties, 
including silicon \cite{no96}, boron nitride \cite{ca16,br19}, 
diamond \cite{ra06,br20}, and graphene \cite{br15,he16,he19}.

In this paper, we employ the path-integral molecular dynamics (PIMD) 
technique to investigate the structural and elastic properties of 
$3C$-SiC across temperatures ranging from $T = 50$ to 1500~K, and 
hydrostatic pressures spanning from $P = -30$~GPa (tension) to 
60~GPa (compression). The interatomic interactions in our simulations 
are described by an effective tight-binding (TB) Hamiltonian.
To assess the precision of the TB outcomes, we have also carried out
DFT calculations at $T = 0$. 
Nuclear quantum effects are assessed by contrasting the findings from 
PIMD simulations with those derived from classical molecular dynamics (MD) 
simulations using the same TB Hamiltonian.
Furthermore, we scrutinize the impact of anharmonicity on the physical 
properties of cubic SiC by comparing our results against those attained 
through a harmonic approximation.
Our findings reveal that quantum corrections induce a significant 
reduction in the elastic constants, bulk modulus, and Poisson's ratio 
at low temperatures. These quantum effects demonstrate their notable 
presence in the structural and elastic properties of $3C$-SiC, 
remaining appreciable at temperatures surpassing 300~K.

Similar path-integral simulations, akin to those presented in this paper, 
have been previously employed to examine nuclear quantum effects in 
carbon-based materials \cite{br15,he16,ra06,br20,he21b}, 
silicon \cite{no96}, and boron nitride \cite{br19,br22}. 
These effects exhibit significance in electronic gaps \cite{ra08} and 
in the isotope dependence of the lattice parameter 
in cubic SiC \cite{he09c}.
Here, we advance the understanding of nuclear quantum effects in 
various properties of materials with cubic symmetry, such as $3C$-SiC. 
In particular, we expand the region of studied pressures to include 
tensile stress, allowing us to investigate this material in a metastable 
region of the $P-T$ phase diagram. This exploration yields information 
on the attractive region of the interatomic potential.
In contrast to earlier simulations of this kind, we present a detailed 
comparison of TB results with those obtained from DFT calculations at 
$T = 0$, providing a more solid foundation for the data derived from 
PIMD simulations.
Moreover, the consideration of SiC permits us to study anharmonicities 
in the quantum vibrational motion of carbon and silicon atoms within 
a binary compound, extending beyond those corresponding to monoatomic 
materials such as diamond and crystalline silicon. This is particularly 
observable in the kinetic energy and mean-square displacement of 
C and Si atoms in $3C$-SiC.

The paper is structured as follows. In Sec.~II, we present the 
computational methods used in the calculations, including the 
tight-binding procedure, path-integral molecular dynamics, and the 
DFT method. In Sec.~III, we outline a harmonic approximation 
for elastic constants and vibrational density of states, 
employed to analyze anharmonicities in our results. 
The internal energy and crystal volume of $3C$-SiC, derived from 
PIMD simulations, are discussed in Secs.~IV and V, respectively. 
Sec.~VI provides results for atomic mean-square displacements. 
Data on the elastic constants and bulk modulus at finite temperatures 
are presented in Secs.~VII and VIII. 
Finally, the main outcomes are summarized in Sec.~IX.

\section{Method of calculation}

\subsection{Tight-binding method}

We investigate nuclear quantum effects on the structural and elastic 
properties of $3C$-SiC, focusing on the quantum delocalization 
of atomic nuclei and its impact on various physical properties of the solid.
For this purpose, two main components are required. Firstly, a suitable 
potential is necessary to define interatomic interactions within 
the material. Such potentials are typically derived from {\em ab-initio} 
techniques, tight-binding Hamiltonians, or empirical models. 
This establishes a Born-Oppenheimer surface for nuclear motion.
Secondly, a method is needed to account for quantum dynamics 
in the configuration space of nuclear coordinates, using the chosen 
interatomic potential. 
This necessitates finite-temperature simulations grounded in quantum 
statistical physics, as opposed to the classical statistics often used 
in atomistic simulations (such as molecular dynamics or Monte Carlo). 
In this paper, we achieve this through PIMD simulations, 
described in Sec.~II.B.

Our simulations are conducted under the adiabatic (Born-Oppenheimer) 
approximation. The potential energy surface for nuclear dynamics is 
derived from an effective tight-binding Hamiltonian \cite{po95}.
While it is possible in principle to employ {\em ab-initio} methods for 
finite-temperature simulations, such an approach would significantly 
limit the duration of simulation trajectories or the feasible system size 
due to computational constraints.
The approach we utilize takes into account both the quantum nature of 
electrons (via the TB Hamiltonian) and atomic nuclei (through the use 
of path integrals). This allows for the direct inclusion of electron-phonon 
and phonon-phonon interactions in our PIMD simulations.

We compute interatomic forces and total energies using the non-orthogonal 
TB Hamiltonian developed by Porezag {\it et al.} \cite{po95}, which is 
grounded in DFT calculations within the local density approximation (LDA).
The specific TB parameterization for structures containing both C and Si 
atoms is detailed in Ref.~\cite{gu96}. In this parameterization, atomic 
orbitals are obtained as eigenfunctions of appropriately constructed 
pseudoatoms, with the valence electron charge density situated near the 
nucleus. The overlap matrices between atomic orbitals and Hamiltonian 
matrix elements are tabulated as functions of internuclear distance. 
Additionally, the short-range repulsive portion of the potential is 
fitted to self-consistent LDA data from relevant reference 
systems \cite{go97}.
The non-orthogonality of the atomic basis plays a pivotal role in 
ensuring the transferability of the TB parametrization to complex 
systems \cite{po95}.

This tight-binding model has previously been applied to investigate 
bulk SiC \cite{sc-me96,sc-be05}, reconstructions of its surfaces 
\cite{gu96}, as well as isotopic and quantum effects in the cubic phase 
\cite{ra08,he09c}. It has also been employed to study various properties 
of recently synthesized silicon carbide monolayers \cite{he22,sc-po23}. 
A comprehensive review of the capabilities of TB procedures in accurately 
describing a range of properties in both molecules and condensed matter 
was provided by Goringe {\it et al.} \cite{go97}.

\begin{figure}
\vspace{-7mm}
\includegraphics[width=7cm]{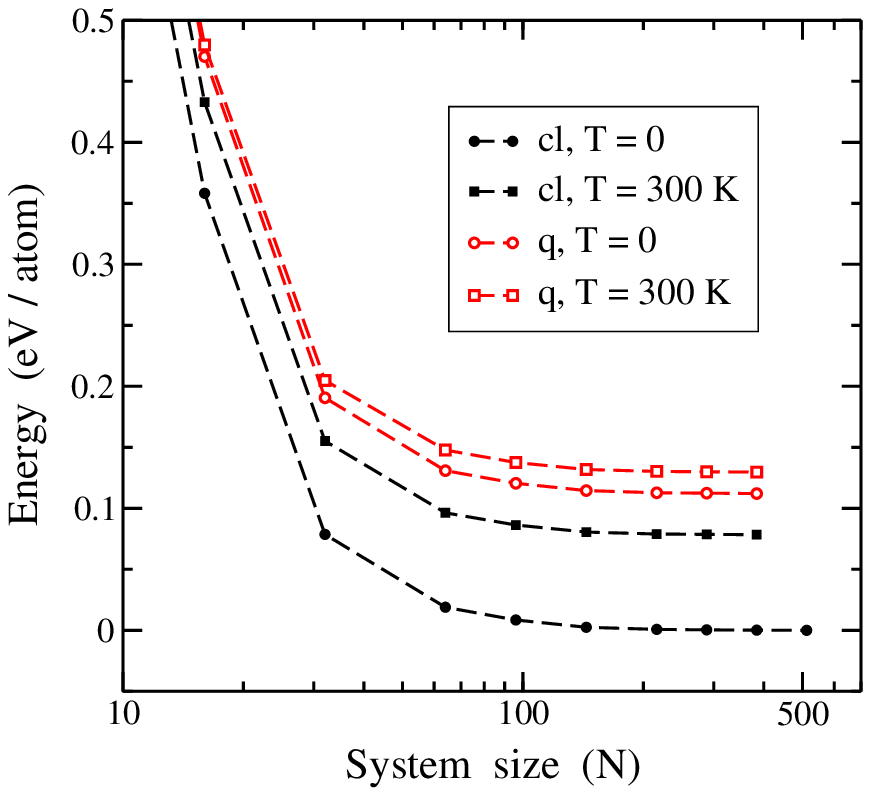}
\vspace{-5mm}
\caption{Energy as a function of system size for $3C$-SiC,
calculated with the TB Hamiltonian used in this work.
Solid circles: $T = 0$; solid squares: classical MD simulations for
$T = 300$~K; open circles: quantum results for $T \to 0$;
open squares: PIMD for $T = 300$~K. Labels ``cl'' and ``q'' refer
to classical and quantum results, respectively.
Lines are guides to the eye.
}
\label{f1}
\end{figure}

To sample the electronic degrees of freedom in reciprocal space, 
we consider only the $\Gamma$ point (${\bf k} = 0$) in 
this paper. Including larger ${\bf k}$ sets in the calculations leads 
to a slight shift in total energy, but with negligible impact on the 
energy differences presented below. This shift results in a minor 
adjustment of the minimum energy $E_0$, which becomes less pronounced 
as the cell size increases. Fig.~1 illustrates this behavior, displaying 
the internal energy of unstressed $3C$-SiC across various cell sizes.
In the figure, solid circles represent the energy attained with the 
TB model ($\Gamma$ point) for the minimum-energy configuration 
($T = 0$, classical). Additionally, we have plotted outcomes for energy 
obtained from classical MD simulations (solid squares) and PIMD 
simulations (open squares) at $T = 300$~K. In both classical and quantum 
cases, we observe a consistent upward shift of the energy relative to 
classical results at $T = 0$. Something similar happens for the
quantum zero-point energy (open squares) derived from extrapolation 
of finite-temperature results of PIMD simulations for different cell 
sizes (see below).

\subsection{Path-integral molecular dynamics}

We investigate the equilibrium properties of $3C$-SiC as functions of 
both temperature and pressure by means of PIMD simulations. 
This technique is rooted in Feynman's path-integral formulation of 
statistical mechanics \cite{fe72}, a valuable nonperturbative approach 
for exploring quantum systems at finite temperatures.
In practical implementations of this computational method, each quantum 
particle (in our context, an atomic nucleus) is represented as a group of 
$N_{\rm Tr}$ (Trotter number) beads. These beads emulate classical particles, 
collectively forming a ring polymer \cite{gi88,ce95}. This representation 
establishes a {\it classical isomorph}, which effectively samples the 
configuration space, providing accurate values for quantum system properties.
For a more comprehensive understanding of this simulation method,
additional details can be found elsewhere \cite{gi88,ce95,he14,ca17}.

We carried out PIMD simulations in the isothermal-isobaric ($NPT$) 
ensemble, employing established algorithms outlined in the literature 
\cite{tu92,tu98,ma99}. Specifically, we made use of staging coordinates 
to define the positions of the beads within the classical isomorph. 
Furthermore, each staging coordinate was coupled to a chain of four 
Nos\'e-Hoover thermostats, maintaining a constant temperature 
throughout the simulations.
Additionally, a chain of four thermostats was linked to the barostat, 
enabling the necessary volume fluctuations to match the targeted 
pressure \cite{tu10,he14}.

The equations of motion were integrated using the reversible 
reference system propagator algorithm (RESPA), which allows to use
different time steps for the integration of fast and slow 
degrees of freedom \cite{ma96}. 
For the dynamics related to interatomic forces, a time step of 
$\Delta t = 1$ fs was employed. Meanwhile, the evolution of fast dynamical 
variables, including harmonic bead interactions and thermostats, was 
computed with a time step of $\delta t = 0.25$ fs.
To calculate the kinetic energy $E_{\rm kin}$, we used the virial 
estimator. This choice is particularly advantageous as it exhibits a 
statistical uncertainty smaller than that of the potential energy, 
especially at high temperatures \cite{he82,tu10}. Further insights into 
this type of PIMD simulations can be found in various literature 
sources \cite{tu98,he06,he16}.

We conducted simulations using $2 \times 2 \times 2$ and 
$3 \times 3 \times 3$ supercells of the face-centered cubic unit cell of 
$3C$-SiC, under periodic boundary conditions. These supercells comprised 
$N =$ 64 and 216 atoms, respectively.
We sampled the configuration space for temperatures ranging from 
50 to 1500~K and pressures spanning from $-30$ to 60~GPa. 
For typical simulation runs, we performed $2 \times 10^5$ PIMD steps for 
system equilibration, followed by $8 \times 10^6$ steps to compute 
average properties.
The Trotter number $N_{\rm Tr}$ was set to vary with temperature according 
to the relation $N_{\rm Tr} T = 6000~{\rm K}$, approximately ensuring 
constant precision for PIMD results across different temperatures 
\cite{he06,he16}.
We have checked that this election of $N_{\rm Tr}$ provides adequate
convergence for our quantum model of $3C$-SiC. In particular, we have
checked this convergence for $T = 300$~K, considering values of the
Trotter number up to $N_{\rm Tr} = 60$, and found results for the
variables considered here which coincide within statistical error bars 
with those found for $N_{\rm Tr} = 20$. Thus, for the energy we
obtained differences smaller than 1 meV/atom.

As the system size $N$ increases, the simulations effectively sample 
vibrational modes with longer wavelengths $\lambda$. In practical terms, 
there exists an effective wavelength cutoff at 
$\lambda_{\rm max} \approx L$, where $L = n a$ (with $a$ being the 
lattice parameter and $n$ equal to 2 or 3 in our case). This corresponds 
to a wavenumber cutoff of 
$k_{\rm min} \approx 2 \pi / L$, where $k = |{\bf k}|$. 
Given that $N \sim L^3$, we observe that $k_{\rm min} \sim N^{-1/3}$.
This wavenumber cutoff, coupled with employing a single ${\bf k}$ point, 
could cause a lack of convergence in calculated magnitudes 
vs system size $N$. We have verified the agreement of results obtained 
for $N = 64$ and 216 atoms, specifically for the energy difference 
$E - E_0$ and the volume $V$, within statistical error bars.

In order to gauge the magnitude of quantum effects calculated from 
our PIMD simulations, we also performed classical MD simulations using 
the same TB Hamiltonian. This corresponds to setting the Trotter number 
to one, which results in the merging of ring polymers into single beads.

An alternative approach for studying anharmonic effects in condensed 
matter involves the use of self-consistent phonon or quasi-harmonic 
approximations, where vibrational mode frequencies are assumed to be 
volume-dependent \cite{de96,sc-ca10b,sc-sh89,sc-ko66}. 
This approach enables the incorporation of temperature effects and 
reveals anharmonicities at $T = 0$ without the need for extrapolation, 
as required by path-integral methods. The quasi-harmonic approximation
has been effectively applied to analyze phenomena such as thermal 
expansion and isotopic effects in solids \cite{mo05,de96,de19b,sc-pe22}, 
as well as the properties of small clusters \cite{sc-sh89}
and molecules \cite{sc-ca10b}.

\subsection{DFT calculations}

To assess the accuracy of the employed TB procedure in describing the 
properties of $3C$-SiC, we carried out state-of-the-art DFT calculations 
for this material. For this purpose, we utilized the Quantum-ESPRESSO 
package for electronic structure calculations \cite{sc-gi09,sc-gi17}.

In particular, we adopted the Perdew-Burke-Ernzerhof 
exchange-correlation functional in its solid-state version (PBEsol) 
\cite{sc-pe08}, along with a plane-wave basis set featuring cutoffs of 
45 Ry for the kinetic energy and 400 Ry for the charge density.
For both C and Si atoms, projector-augmented-wave (PAW) 
pseudopotentials were employed \cite{sc-ps23}.

We considered a cubic zinc-blende structure cell of SiC containing 
8 atoms subject to periodic boundary conditions.
For Brillouin zone integration, we employed a $10 \times 10 \times 10$ 
Monkhorst-Pack grid \cite{sc-mo76}.

{\em Ab-initio} electronic-structure calculations have been previously 
carried out to explore various properties of cubic silicon carbide. 
These investigations encompass lattice-dynamical, structural, mechanical, 
electronic, and thermodynamic properties 
\cite{sc-ch86,sc-pa94b,sc-ka94,sc-ka94b,sc-ca20}.
Such calculations have proven highly valuable for investigating the phase 
diagram of silicon carbide, particularly regarding phase transitions 
under high pressure \cite{sc-ch87,sc-le15b,sc-ki17,sc-sh18,sc-ra21}.

\section{Harmonic approximation}

To asses the relevance of anharmonicity in the results of our PIMD 
simulations for $3C$-SiC, we consider a harmonic approximation (HA) 
for the atomic vibrational modes. While this approximation is generally
reliable at low $T$ in solids, anharmonicity typically becomes
more pronounced as $T$ increases. This leads to a progressive
deviation of the harmonic approach from the more accurate
atomistic simulations.
Within the HA, frequencies are treated as independent 
of temperature, thereby excluding considerations of volume changes 
(thermal expansion).

To establish a reference for the subsequent analysis of thermal and 
nuclear quantum effects, we assess the elastic stiffness constants, 
$C_{ij}$, in the classical low-temperature limit. 
 These elastic constants for $3C$-SiC at $T = 0$ 
have been derived from the harmonic dispersion relation of acoustic 
phonons.  This calculation involved diagonalization of the dynamical 
matrix obtained from the TB Hamiltonian \cite{sc-he23}.

Sound velocities in a solid can be derived by assessing the slope of 
acoustic phonon branches close to the $\Gamma$ point. 
Specifically, we obtain these velocities by evaluating the derivative 
$\partial \omega / \partial k$ along high symmetry directions of the 
Brillouin zone. In the context of cubic crystals, this relation is 
established as \cite{ki05,yu96}:
\begin{equation}
  C_{11} = \rho \left( \frac{\partial \omega_{\rm LA}}{\partial k_x }
           \right)^2_{\Gamma}  \; ,
\label{c11}
\end{equation}
for the longitudinal acoustic (LA) band along the $[100]$ direction, and
\begin{equation}
  C_{12} = C_{11} - 2 \rho \left( \frac{\partial \omega_{\rm TA_2}}
           {\partial k}  \right)^2_{\Gamma}  \; ,
\label{c12}
\end{equation}
\begin{equation}
  C_{44} =  \rho \left( \frac{\partial \omega_{\rm TA_1}}
           {\partial k}  \right)^2_{\Gamma}  \; ,
\label{c44}
\end{equation}
for the  transverse acoustic bands, TA$_1$ and TA$_2$, along the 
$[110]$ direction. 
In these equations, $\rho$ denotes the solid's density.
For the HA, our calculations yield the following elastic constants: 
$C_{11} = 452.9$~GPa, $C_{12} = 141.1$~GPa, and $C_{44} = 246.7$~GPa. 
We have confirmed that these values align with those obtained 
for the different phonon bands along other symmetry directions
in ${\bf k}$-space.

The isothermal bulk modulus of cubic crystals can be determined from 
the elastic constants through the expression \cite{ki05}:
\begin{equation}
   B = \frac13 (C_{11} + 2 C_{12})  \; .
\label{bulkm}
\end{equation}
From this formula, we find a bulk modulus $B_0$ = 245.0 GPa for the
minimum-energy configuration (classical minimum).

\begin{figure}
\vspace{-16mm}
\includegraphics[width=8cm]{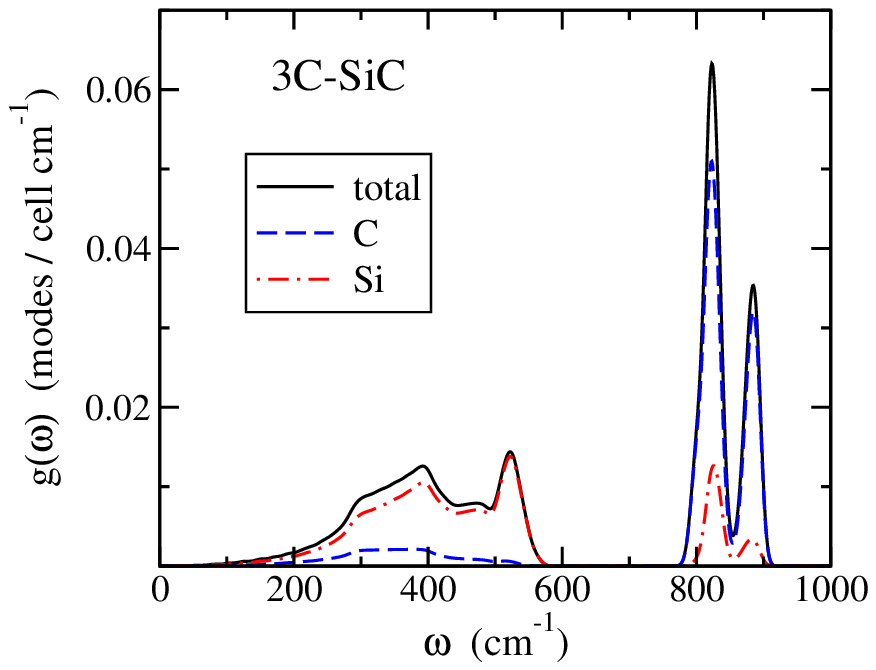}
\vspace{-5mm}
\caption{Vibrational density of states of $3C$ silicon carbide,
calculated in the HA for the TB Hamiltonian used in this work.
Dashed and dashed-dotted curves depict the VDOS
corresponding to carbon, $g_{\rm C}(\omega)$, and
silicon, $g_{\rm Si}(\omega)$, respectively.
The solid line represents the total density of states,
$g(\omega)$.
}
\label{f2}
\end{figure}

To directly assess the anharmonicity of atomic vibrations, 
we have calculated the vibrational density of states (VDOS) for 
the entire Brillouin zone within the HA. This computation 
was executed via numerical integration, following the method 
outlined in Ref.~\cite{ra86}.
In Fig.~2, we present the resultant VDOS for $3C$-SiC, with dashed 
and dashed-dotted lines denoting the respective contributions from carbon 
and silicon. The solid line represents the cumulative VDOS encompassing 
both constituents. We will use the notations 
$g_{\rm C}(\omega)$ and $g_{\rm Si}(\omega)$ to call the VDOS 
contributions from C and Si atoms, while the overall VDOS is expressed 
as $g(\omega) = g_{\rm C}(\omega) + g_{\rm Si}(\omega)$. 

The quantum-mechanical vibrational energy per atom at temperature $T$ 
within the HA is given by:
\begin{equation}
 E_{\rm vib}  =  \frac{1}{2N} \sum_{r,\bf k} \hbar \, \omega_r({\bf k})
   \coth \left( \frac12 \beta \hbar \, \omega_r({\bf k}) \right)  \, ,
\label{evib}
\end{equation}
where $\beta = 1 / k_B T$ (with $k_B$ being Boltzmann's constant). 
The index $r$ ($r$ = 1, ..., 6) designates the phonon branches, and 
the summation over wavevectors ${\bf k}$ traverses the Brillouin zone.
Alternatively, based on the continuous approximation of the VDOS 
$g(\omega)$, the energy $E_{\rm vib}$ can be calculated as:
\begin{equation}
 E_{\rm vib} = \frac14 \int_0^{\omega_{max}} \hbar \, \omega
     \coth \left( \frac12 \beta \hbar \, \omega \right)
      g(\omega) d \omega  \, ,
\label{evib2}
\end{equation}
where $\omega_{max}$ stands for the maximum frequency in the phonon 
spectrum.  Note that we use the normalization condition 
\begin{equation}
  \int_0^{\omega_{max}}  g(\omega)  d \omega  = 6,
\label{intg}
\end{equation}
to account for the six degrees of freedom present in a crystallographic 
unit cell (comprising one C and one Si atom).

\section{Energy}

In this section, we present and analyze the internal energy of 
$3C$-SiC, extracted from PIMD simulations conducted within the $NPT$ 
ensemble at various pressures and temperatures. This simulation approach 
provides distinct evaluations of the potential and kinetic energy
of the system \cite{ra11,he14,he22}. This enables the exploration of 
lattice vibration anharmonicities by scrutinizing differences between 
both energies, which coincide for harmonic systems. 

\begin{figure}
\vspace{-7mm}
\includegraphics[width=7cm]{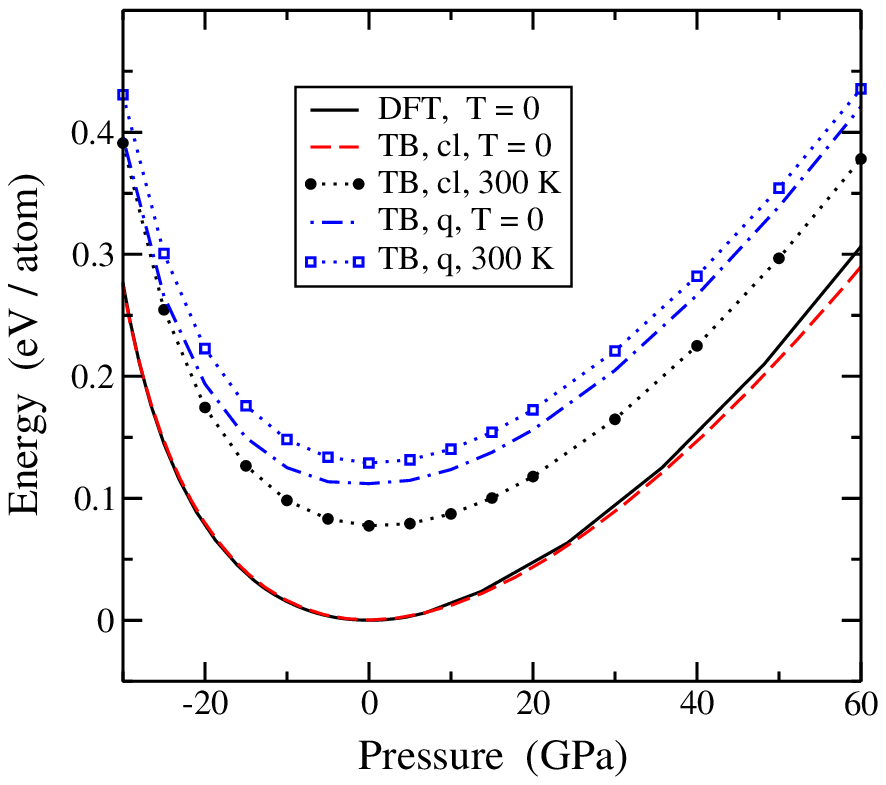}
\vspace{-5mm}
\caption{Energy per atom vs hydrostatic pressure.
The solid and dashed lines represent results obtained
from DFT and TB calculations, respectively, at $T = 0$.
Symbols depict results from classical MD (solid circles)
and PIMD simulations (open squares) at $T = 300$~K.
The dashed-dotted line represents the quantum limit for
$T \to 0$.
Labels ``cl'' and ``q'' refer to classical and quantum
results, respectively.
Dotted lines are guides to the eye.
}
\label{f3}
\end{figure}

For a specified combination of pressure and temperature, the internal 
energy can be expressed as $E = E_0 + E_{\rm pot} + E_{\rm kin}$, where 
$E_{\rm pot}$ and $E_{\rm kin}$ correspond to the potential and kinetic 
energy, respectively. $E_0$ denotes the energy of the classical model 
at zero temperature (minimum-energy configuration).  Fig.~3 displays 
the energy difference $E - E_0$ as a function of hydrostatic pressure. 
The solid and dashed curves portray the energy outcomes derived 
at $T = 0$ via DFT and TB calculations, respectively. The energy 
reference is established at the unstressed material state ($P = 0$).
The energy curves closely align in the pressure range 
spanning from a tensile pressure of $P = -30$~GPa to a compressive 
pressure of approximately 40~GPa. 
Beyond this interval of compressive stress, a divergence becomes
apparent between the TB and DFT energy curves, the former exhibiting 
a slower growth rate compared to the latter.

In Fig.~3, symbols are employed to represent the outcomes of 
simulations performed at $T = 300$~K. Solid circles correspond to classical 
MD results, while open squares represent PIMD simulations.
The classical results exhibit a nearly uniform increase of $3 k_B T$ relative 
to the TB results at $T = 0$. This signifies the contribution of thermal 
energy per atom at the given temperature. In the case of PIMD simulations, 
we observe that at a pressure of $P = 0$, there is an enhancement of 
51~meV/atom compared to the classical data. This difference in energy grows 
from 40 to 57~meV/atom within the pressure range spanning from $-30$ to 60~GPa.
The average phonon frequency $\overline{\omega}$ escalates with increasing 
pressure, leading to an amplification of the energy difference $\delta E$ 
between quantum and classical models, especially at relatively low 
temperatures, including 300~K. 
The dashed-dotted line in this figure represents the quantum limit of 
the energy for $T \to 0$, extrapolated from our finite-temperature PIMD 
simulations in the plotted pressure range.
In the low-$T$ limit, the rate of change 
$\partial (\delta E) / \partial P$ is given by
$\frac32 \hbar \, \partial \overline{\omega}/ \partial P > 0$. 

\begin{figure}
\vspace{-7mm}
\includegraphics[width=7cm]{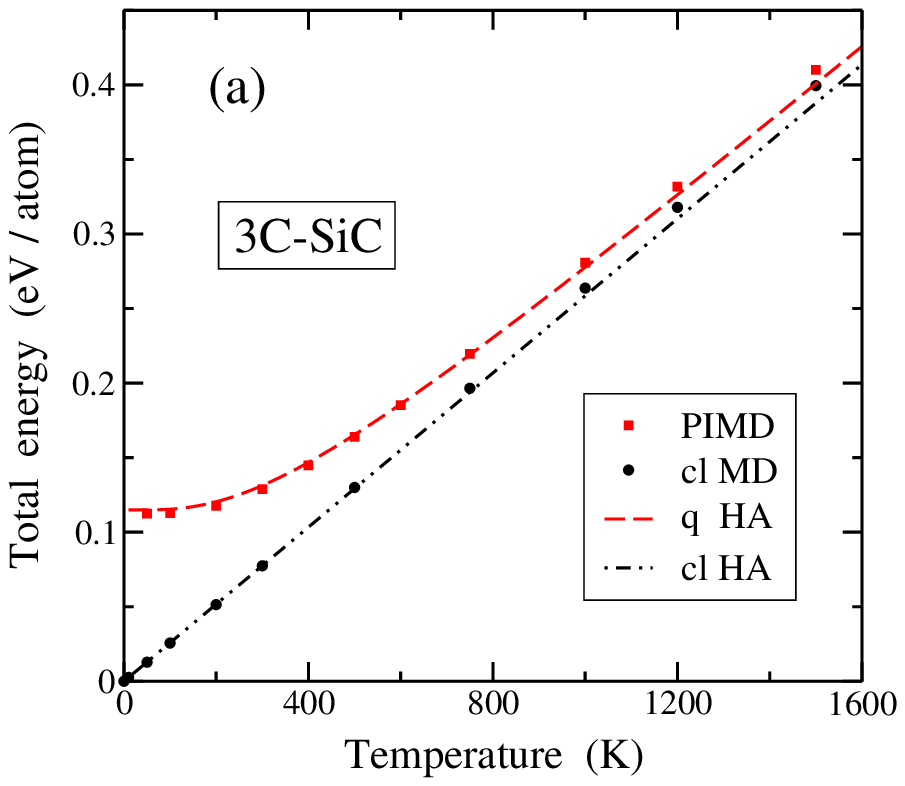}
\includegraphics[width=7cm]{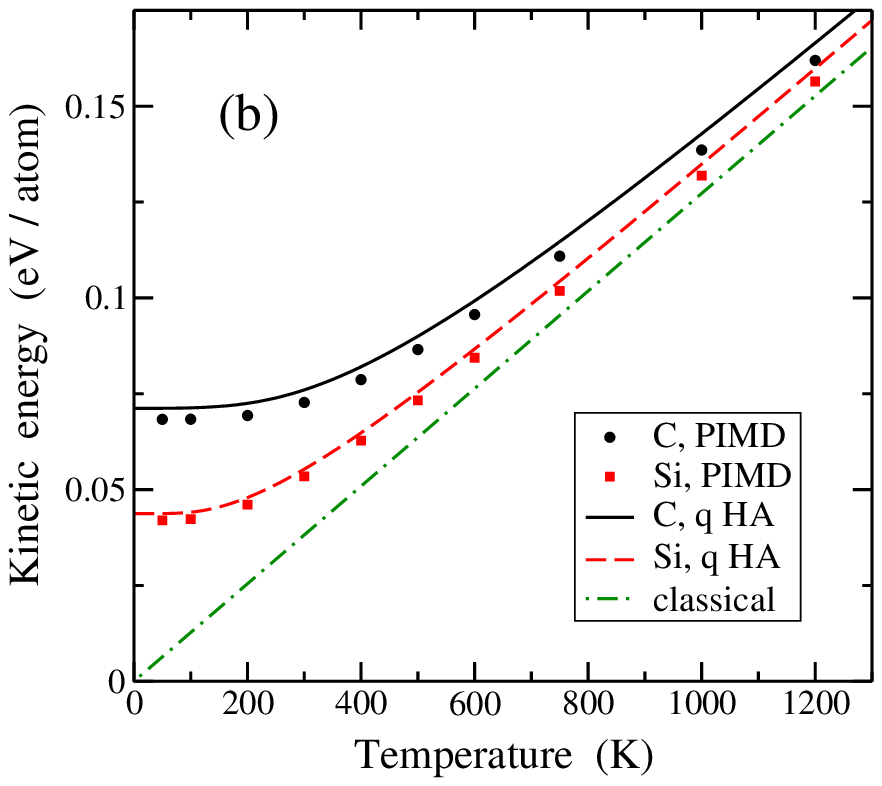}
\vspace{-5mm}
\caption{(a) Energy per atom vs temperature.
Symbols display results of classical MD (circles) and
PIMD simulations (squares.)
The dashed line represents the HA obtained from the VDOS
using Eq.~(\ref{evib2}). The dashed-dotted line depicts
the classical thermal energy per atom: $E^{\rm cl} = 3 k_B T$.
Labels ``q'' and ``cl'' refer to quantum and classical results,
respectively.
(b) Temperature dependence of the kinetic energy of C (solid
circles) and Si atoms (solid squares), obtained from PIMD simulations.
Solid and dashed lines represent the kinetic energy found
from the corresponding VDOS, $g_{\rm C}(\omega)$ and
$g_{\rm Si}(\omega)$, respectively.
The dashed-dotted line illustrates the classical kinetic energy
$E_{\rm kin}^{\rm cl} = 3 k_B T / 2$.
}
\label{f4}
\end{figure}

In Fig.~4(a), we present the temperature dependence of the internal energy, 
$E - E_0$, which was obtained from PIMD simulations of cubic SiC at zero 
pressure, indicated by solid squares. Additionally, we display 
the internal energy values obtained from classical MD simulations, 
represented by solid circles.
As temperature decreases, the quantum results converge to an energy 
$E = E_0 + E_{\rm ZP}$, where $E_{\rm ZP}$ is the zero-point energy, which 
results to be 112~meV/atom. At low temperatures, classical simulations exhibit 
a dependence described by $E - E_0 \propto T$, in line with predictions 
derived from the equipartition principle in classical statistical mechanics 
for harmonic vibrations: $E - E_0 = 3 k_B T$.
At high temperature, classical MD simulations show slight deviations from 
this linear dependence due to the anharmonicity of lattice vibrations. 
Notably, the zero-point energy obtained from PIMD simulations is on the order 
of the classical thermal energy at 450~K, which corresponds to approximately 
one third of the material's Debye temperature \cite{sc-zy96}.
As the temperature increases, the results of both PIMD and classical MD 
simulations gradually converge. However, even at $T = 1000$~K, a difference
of 17~meV/atom remains between both sets of data.

In Fig.~4(b), the kinetic energy as a function of temperature is presented. 
The symbols represent data from our PIMD simulations: circles 
indicate C atoms, while squares denote Si atoms. The solid and dashed 
lines on the graph represent the results obtained from the HA for C and Si, 
respectively.
It is noticeable that in both cases, the symbols lie below their corresponding 
lines. This observation indicates that anharmonicity leads to a reduction in 
kinetic energy. According to the outcomes of quantum simulations at low 
temperatures, the kinetic energy for carbon and silicon atoms is 68.3 and 
42.0 meV, respectively. This yields a ratio of 
$E_{\rm kin}^{\rm C} / E_{\rm kin}^{\rm Si}$ = 1.63 as the temperature 
approaches absolute zero.
Comparing these low-temperature results to those provided by the HA, 
it becomes apparent that, due to anharmonicity, there is a reduction of 
4\% in each case.  The dashed-dotted line featured in Fig.~4(b) 
represents the classical kinetic energy per atom, which remains constant 
regardless of the atomic mass: $E_{\rm kin} = 3 k_B T / 2$.

\begin{figure}
\vspace{-7mm}
\includegraphics[width=7cm]{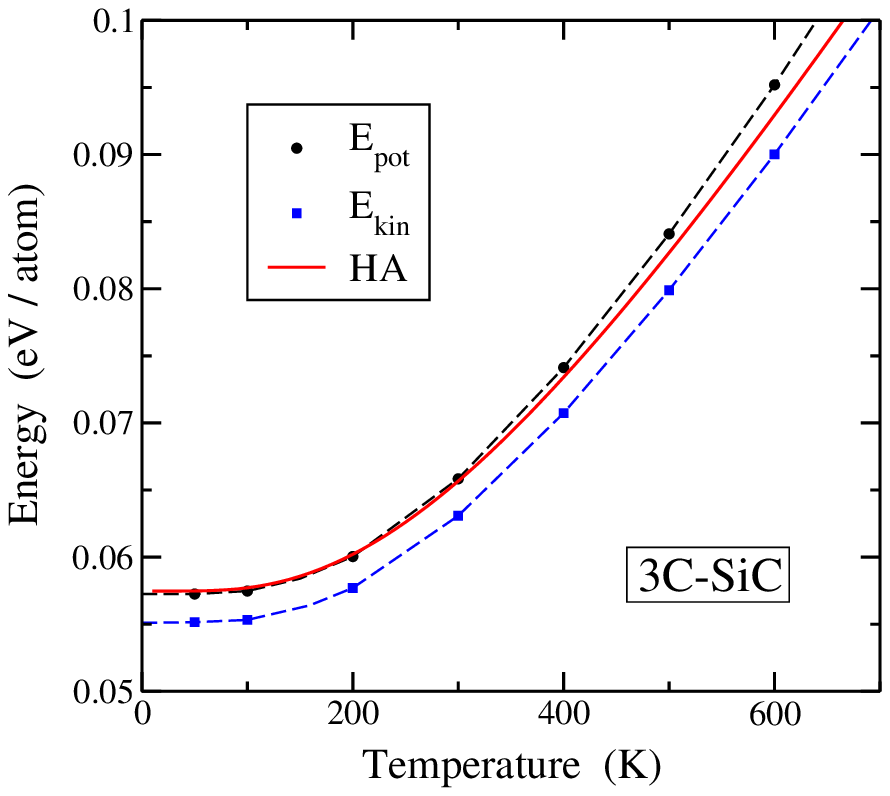}
\vspace{-5mm}
\caption{Mean kinetic ($E_{\rm kin}$, squares) and potential energy
($E_{\rm pot}$, circles) per atom vs temperature, as derived
from PIMD simulations.
The solid line represents the harmonic expectancy obtained from
the VDOS as $E_{\rm kin} = E_{\rm pot} = E_{\rm vib} / 2$, using
Eq.~(\ref{evib2}). Dashed lines are guides to the eye.
}
\label{f5}
\end{figure}

An evaluation of lattice-vibration anharmonicity can be achieved by 
comparing the potential and kinetic energy derived from PIMD simulations. 
In the context of harmonic vibrations, the expectation is that 
$E_{\rm pot} = E_{\rm kin}$ (as derived from the virial theorem), 
so that any deviation of the ratio $E_{\rm kin} / E_{\rm pot}$ from
unity is an indication of anharmonicity.
In Fig.~5, we present the temperature-dependent behavior of both overall 
kinetic and potential energy. As before, the symbols on the graph 
correspond to data obtained from PIMD simulations, while the solid 
line represents the outcome for the harmonic approximation.
At low temperatures, a ratio of 
$E_{\rm pot} / E_{\rm kin} = 1.04$ is obtained. This ratio displays 
an increase as the temperature rises, reaching a value of 1.08 at 1000~K.

At low temperature, an apparent difference emerges between the potential 
and kinetic energy. The potential energy tends to closely align with 
the harmonic expectation, in contrast to the behavior observed for
the kinetic energy, as discussed earlier for C and Si atoms.
This fact has also been noted in quantum simulations of various other 
materials and for impurities in crystalline solids \cite{he16,he95}. 
The reason behind this can be comprehended by examining energy 
shifts obtained through time-independent perturbation methods.
To illustrate, in the case of a one-dimensional perturbed harmonic 
oscillator, where the potential energy is given by 
$V(x) = m \omega^2 x^2 / 2$ and a perturbation $W(x) = A x^3 + B x^4$ 
is introduced, the first-order variation in the ground-state energy is 
attributed to a change in kinetic energy, while the potential energy 
remains unaffected, as in the unperturbed oscillator \cite{he95,la65}.
For a more detailed explanation, we refer to Appendix~A.

\section{Volume}

In the case of unstressed $3C$-SiC, the volume of the crystal obtained 
from classical MD simulations displays an almost linear dependence on 
temperature. The slope $\partial V/ \partial T$ of this relationship 
increases gradually as the temperature rises. At low $T$, 
the volume converges toward a value of $V_0 = 10.277$~\AA$^3$/atom, 
corresponding to the minimum-energy volume (lattice parameter 
$a_0 = 4.348$~\AA).
For the purpose of comparison, our DFT calculations yield a minimum-energy 
volume of $V_0 = 10.346$~\AA$^3$/atom (lattice parameter $a_0 = 4.358$~\AA). 
This value is close to the results obtained from previous {\em ab-initio} 
studies \cite{sc-le15b}.

For each given temperature $T$, the volume obtained through quantum 
PIMD simulations consistently exceeds the corresponding classical value. 
This trend converges toward a volume of $V_{\rm min} = 10.355$~\AA$^3$/atom 
as the temperature approaches absolute zero ($a_{\rm min} = 4.359$~\AA). 
Consequently, a zero-point volume expansion of 0.8\% in comparison to 
the classical minimum is observed, translating to a lattice parameter 
increase of $\delta a = 1.1 \times 10^{-2}$~\AA.
The simulation results from both the quantum and classical approaches 
gradually align as the temperature increases. At 300 K, the difference 
between the two sets of data is approximately half of its value at 
the low-temperature limit.
Remarkably, the outcomes of PIMD simulations employing the TB Hamiltonian 
closely approximate the experimental data derived from x-ray diffraction 
of cubic SiC under ambient conditions. Specifically, at 297~K the 
experimental lattice parameter was found to be 
$a = 4.36$~\AA\ \cite{sc-sl75,ra08}.

The quantum zero-point dilation in a solid is governed by the presence of 
anharmonicities in its vibrational modes. This has predominantly been 
explored within a quasi-harmonic approximation. In this
approach, each individual phonon mode contributes to the $T = 0$
dilation by the product of its zero-point energy and the associated 
Gr\"uneisen parameter $\gamma_{\omega}$ \cite{mo05,de96,he20c}.
An alternative approach consists in defining an overall parameter, 
denoted as $\overline{\gamma}$, derived from the average frequency 
$\overline{\omega}$ as follows:
\begin{equation}
 \overline{\gamma} = - \frac {\partial ({\rm log} \, \overline{\omega}) }
          {\partial ({\rm log} \, V) }  \; .
\label{gamma}
\end{equation}
This formulation proves valuable in investigating the influence of 
anharmonicity on the thermodynamic properties of solids \cite{as76}.

At $T = 0$, the energy $E$ for a specific volume $V$ can be written
as the sum of two components: $E = E_{\rm cl} + E_{\rm ZP}$. Here, 
$E_{\rm cl}$ corresponds to the classical energy as depicted in Fig.~3, 
and $E_{\rm ZP} = 3 \hbar \overline{\omega} / 2$ represents the 
zero-point energy per atom. By employing the equilibrium condition 
under zero pressure, $\partial E / \partial V = 0$, and referring to 
Appendix~B for more details, we derive the equation:
\begin{equation}
  \left( \frac{\partial E_{\rm cl}} {\partial V} \right)_{V_{\rm min}}
   \approx  \frac{\overline{\gamma} E_{\rm ZP}} {V_0} \; .
\label{dev}
\end{equation}
In this context, $V_{\rm min}$ signifies the volume corresponding to 
the quantum ground state. Considering the volume dependence of 
$E_{\rm cl}$, as given by the TB Hamiltonian, we determine the left-hand 
side of Eq.~(\ref{dev}) to be 1.82 GPa. This computation leads to 
a value of $\overline{\gamma}$ equal to 1.05, consistent with the 
Gr\"uneisen parameter values available in the literature for $3C$-SiC 
\cite{sc-va15,sc-ch87}.
As a result, the quantum zero-point dilation can be expressed using 
Eq.~(\ref{vmin2}) as follows:
\begin{equation}
 V_{\rm min} - V_0 = \frac{ \overline{\gamma} E_{\rm ZP} } {B_0} \; .
\label{vmv0}
\end{equation}
Given a value of $\overline{\gamma}$ equal to 1.05, Eq.~(\ref{vmv0}) 
yields a volume of $V_{\rm min} = 10.354$ \AA$^3$/atom, which is 
consistent with the low-temperature volume directly obtained from 
PIMD simulations.

\begin{figure}
\vspace{-7mm}
\includegraphics[width=7cm]{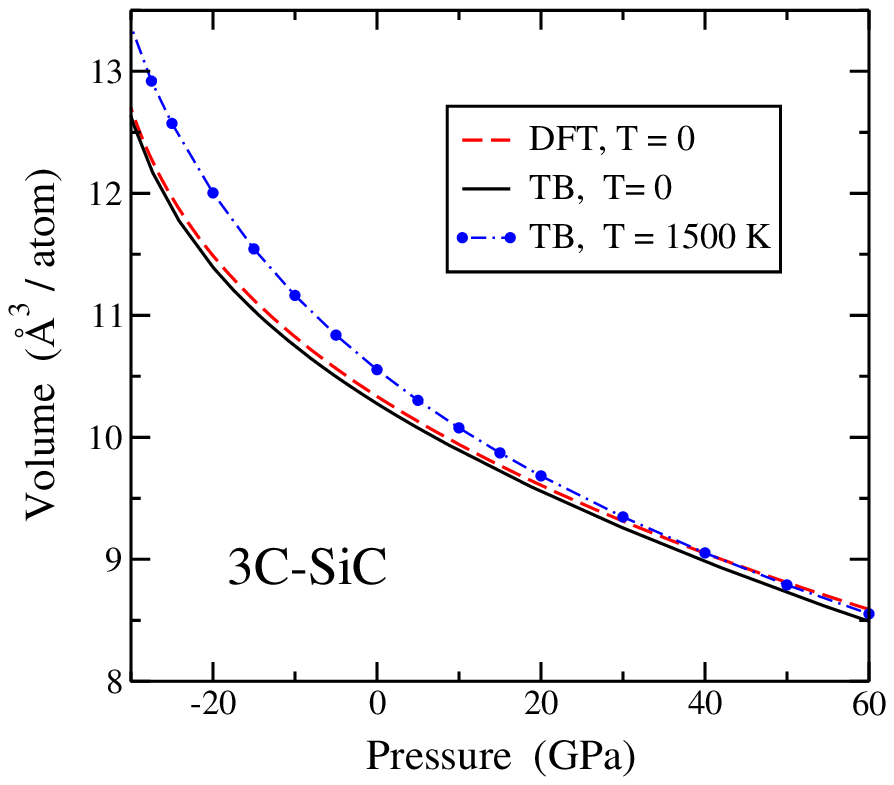}
\vspace{-5mm}
\caption{Volume vs pressure, as obtained from DFT (dashed curve)
and tight-binding calculations (solid curve) at $T = 0$.
Circles represent results of classical TB MD simulations
at 1500~K. Error bars are in the order of the symbol size.
The dashed-dotted line is a guide to the eye.
}
\label{f6}
\end{figure}

We will now analyze the influence of hydrostatic pressure on the crystal 
volume $V$. In Fig. 6, we display the pressure-induced changes in volume 
per atom at $T = 0$, as determined through DFT calculations (depicted 
by the dashed curve) and TB calculations (represented by the solid curve). 
These two curves closely align within the pressure range up to approximately 
30~GPa, although the DFT result is slightly higher than the TB data. 
As pressure increases further, the disparity between both datasets 
gradually becomes more pronounced.
The solid circles featured in Fig.~6 denote data points derived from 
classical MD simulations at $T = 1500$~K. A comparison between these 
results and the TB calculations at $T = 0$ reveals a notable thermal 
expansion under tensile pressure. Specifically, this expansion amounts to 
$\delta V$ = 0.277 and 0.604 \AA$^3$/atom for pressures of $P = 0$ and 
$-20$~GPa, respectively. In contrast, the volume exhibits a lesser dilation 
under compression. For instance, at a pressure of $P = 40$~GPa and a 
temperature of 1500~K, the volume expansion is 
$\delta V = 0.067$ \AA$^3$/atom.

Results of our PIMD simulations at low temperatures indicate that
the volume expansion associated to quantum zero-point motion is also
significantly affected by hydrostatic pressure. In fact, for 
increasing compressive stress, the zero-point volume increase
is drastically reduced in comparison to unstressed $3C$-SiC.  
Specifically, for $P = 50$~GPa, the volume change $\delta V$ as $T$ 
approaches zero is calculated to be $7 \times 10^{-3}$ \AA$^3$/atom. 
This is in contrast to the low-temperature expansion of 
$8 \times 10^{-2}$~\AA$^3$/atom observed for cubic SiC at $P = 0$.

The change in volume caused by an applied hydrostatic pressure is linked 
to the material's bulk modulus $B$, as discussed in Sec.~VIII. 
According to our classical data at $T = 0$, the rate of volume change, 
$\partial V / \partial P$, turns out to be 
$-4.19 \times 10^{-2}$ \AA$^3$/(atom GPa) at $P \to 0$. 
In contrast, at a temperature of $T$ = 1000~K, this derivative 
amounts to $-5.32 \times 10^{-2}$~\AA$^3$/(atom GPa).

For a cubic crystal subjected to uniaxial pressure (say $\tau_{xx}$), 
the alteration in volume can be determined using the elastic compliance 
constants \cite{ra20b,ki05,yu96}:
\begin{equation}
 \frac{\delta V}{V} = e_{xx} + e_{yy} + e_{zz} =
            (S_{11} + 2 S_{12}) \tau_{xx}  \; ,
\end{equation}
Furthermore, we can derive the stress derivative of volume as follows:
\begin{equation}
 \frac{\partial V}{\partial \tau_{xx}} = (S_{11} + 2 S_{12}) V 
                 = \frac{V}{3 B}   \; ,
\end{equation}
which provides an alternative formula for computing the bulk modulus  
through classical MD and PIMD simulations. This approach serves as a 
method of cross-validation, thereby enhancing the reliability assessment 
of our methodology.

\section{Atomic mean-square displacements}

In this section, we present an analysis of the atomic mean-square 
displacement (MSD) in $3C$-SiC, spanning a broad temperature region. 
The interplay between quantum motion and the anharmonicity of vibrational 
modes gives rise to discernible effects on the structural and mechanical 
characteristics of solids, particularly at low temperatures.
PIMD simulations offer a means to compute atomic MSDs across varying 
temperatures. These MSDs encompass both a classical (thermal) part
and an intrinsic quantum contribution. The former corresponds to 
the motion of the center of gravity (centroid) of the quantum paths 
associated with the atomic nuclei. In contrast, the latter is linked 
to the average size of the ring polymers that depict the quantum behavior 
of the nuclei.

\begin{figure}
\vspace{-7mm}
\includegraphics[width=7cm]{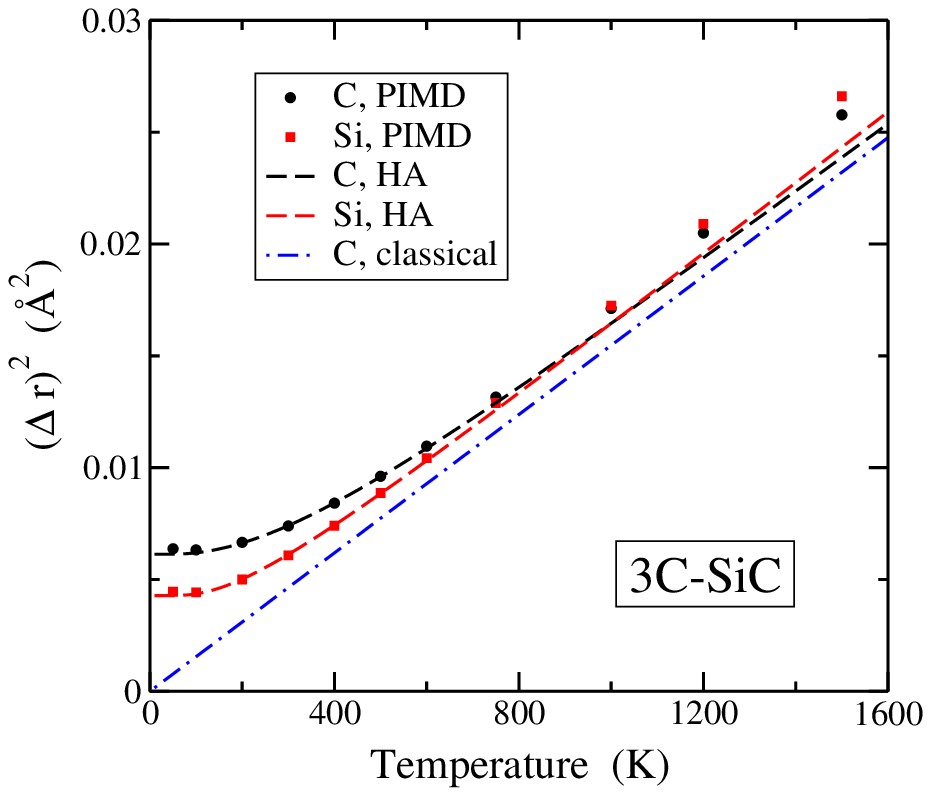}
\vspace{-5mm}
\caption{Atomic mean-square displacement, $(\Delta {\bf r})^2$,
in $3C$-SiC as a function of temperature. Solid symbols represent
results of PIMD simulations for carbon (circles) and silicon
(squares). Error bars are in the order of the symbol size.
Dashed lines correspond to MSDs for C and Si obtained from
the respective VDOS using Eq.~(\ref{dr2c}).
The dashed-dotted line represents the carbon MSD obtained in
the classical approximation by means of Eq.~(\ref{dr2c2}).
}
\label{f7}
\end{figure}

In Fig.~7, we display the temperature-dependent behavior of the atomic 
MSD for C, represented by circles, and Si, represented by squares. 
These MSD values are derived from PIMD simulations.
For $T \to 0$, we determine the squared displacements $(\Delta {\bf r})^2$ 
to be $6.3 \times 10^{-3}$~\AA$^2$ for C and 
$4.4 \times 10^{-3}$~\AA$^2$ for Si atoms,
which correspond to zero-point motion. The quotient between these 
values is close to the inverse square root of the mass ratio, i.e.,
$(\Delta {\bf r})_{\rm C}^2 / (\Delta {\bf r})_{\rm Si}^2$
$\approx (M_{\rm Si} / M_{\rm C})^{1/2}$.
In the presence of identical effective (harmonic) potentials for
both species, these two ratios should strictly coincide.
The differing environments of C and Si atoms lead to a difference 
in these ratios.

The dashed lines depicted in Fig.~7 were generated using the VDOS 
for C and Si presented in Fig.~2, using the HA. For carbon  atoms, 
this MSD is obtained from the expression:
\begin{equation}
   (\Delta {\bf r})^2_{\rm C} =  \int_{\omega_0}^{\omega_{max}}
        \frac{\hbar} {2 \omega  M_{\rm C}}
     \coth \left( \frac12 \beta \hbar \, \omega \right)
      g_{\rm C}(\omega) d \omega  \, .
\label{dr2c}
\end{equation}
A similar expression is utilized for Si, taking into account the 
atomic mass $M_{\rm Si}$ and the function $g_{\rm Si}$.
At low $T$, the outcomes of quantum simulations closely align with those 
yielded by the HA. As the temperature increases, the simulation data
progressively surpass the HA predictions. Notably, at approximately 
$T = 1000$~K, both the simulation results (represented by symbols) and 
the HA (shown by dashed lines) indicate that $(\Delta {\bf r})^2_{\rm Si}$ 
exceeds $(\Delta {\bf r})^2_{\rm C}$.

The dashed-dotted line in Fig.~7 indicates the MSD of C atoms 
obtained through a classical HA. In this approximation, one has:
\begin{equation}
   (\Delta {\bf r})^2_{\rm C} =  \int_{\omega_0}^{\omega_{max}}
        \frac{k_B T} {\omega^2 M_{\rm C}}
         g_{\rm C}(\omega) d \omega  \, .
\label{dr2c2}
\end{equation}
MSDs obtained from classical calculations are not contingent on atomic mass, 
but rather they vary in accordance with the effective potential experienced
by atomic nuclei. In this specific instance, the interatomic potential 
experienced by C atoms is relatively ``stiffer'' compared to that associated 
with Si atoms. As a result, the classical MSD for silicon becomes larger 
than that of carbon due to the inherent differences in their effective 
potentials. This distinction is the reason for the observation
that $(\Delta {\bf r})^2_{\rm Si} > (\Delta {\bf r})^2_{\rm C}$ in the 
quantum calculations at elevated temperatures, as illustrated in Fig.~7.

One way to evaluate the consistency of results from our PIMD simulations,
both in position and momentum domains, consists in examining the MSDs
$(\Delta {\bf r})^2$ and kinetic energy.
In accordance with Heisenberg's uncertainty principle, the root-mean-square 
deviations for the coordinate $x$ and the corresponding momentum $p_x$ of 
a quantum particle must satisfy the inequality 
$\Delta x \, \Delta p_x \geq \hbar / 2$ (as discussed in, for instance, 
Ref.~\cite{co20a}). This relationship can also be expressed as:
\begin{equation}
   (\Delta p_x)^2  \geq  \frac{\hbar^2}{4 (\Delta x)^2}  \, .
\label{dpx2}
\end{equation}
Similar relations hold for the coordinates $y$ and $z$. 

\begin{figure}
\vspace{-7mm}
\includegraphics[width=7cm]{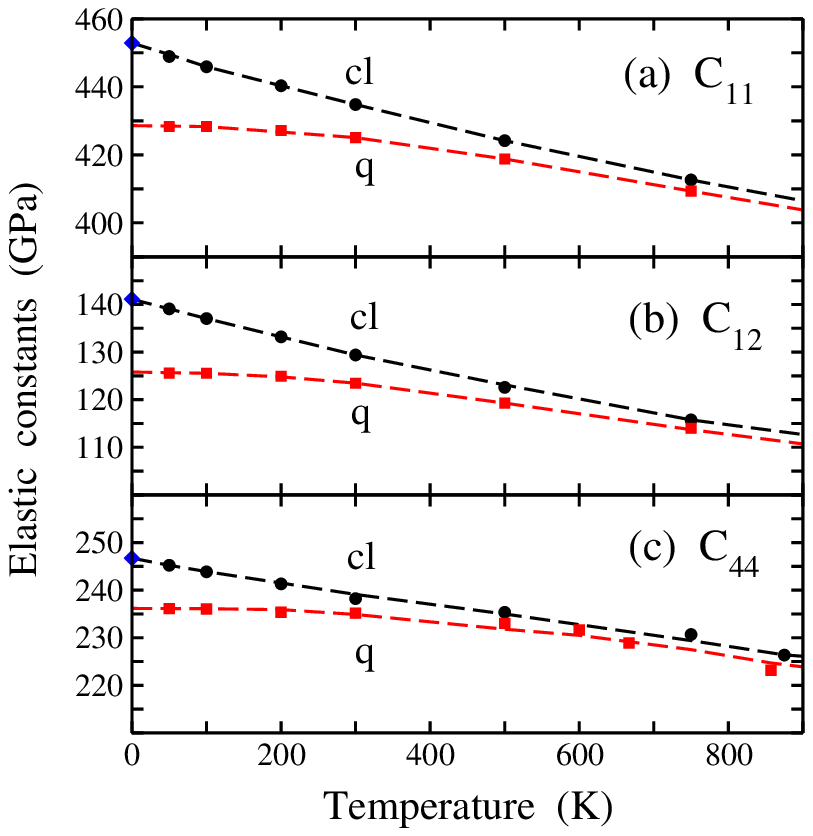}
\vspace{-5mm}
\caption{Temperature dependence of the elastic stiffness constants
of cubic silicon carbide:
(a) $C_{11}$, (b) $C_{12}$, and (c) $C_{44}$, as derived from classical
MD (circles, labeled ``cl'') and PIMD simulations (squares, labeled ``q'').
Error bars are in the order of the symbol size.
Solid diamonds represent in each case the classical value at $T = 0$,
calculated from the phonon dispersion bands.
Dashed lines are guides to the eye.
}
\label{f8}
\end{figure}

In our specific case, the average momentum is zero 
($\langle {\bf p} \rangle = 0$), as usually happens in solids.  
Thus, for an atomic nucleus with mass $M$, the kinetic energy can 
be expressed as:
\begin{equation}
  E_{\rm kin} = \frac{\langle {\bf p}^2 \rangle}{2 M} =
        \frac{(\Delta {\bf p})^2}{2 M}  \; .
\label{ekin}
\end{equation}
For cubic crystal structures, where 
$(\Delta x)^2 = (\Delta y)^2 = (\Delta z)^2$, combining 
Eqs.~(\ref{dpx2}) and (\ref{ekin}) yields:
\begin{equation}
  E_{\rm kin}  \geq  \Lambda  \equiv
        \frac{9 \hbar^2}{8 M (\Delta {\bf r})^2 }  \; ,
\label{ekinf}
\end{equation}
where $\Lambda$ is a function of the atomic MSD. This inequality 
establishes a lower limit for the kinetic energy, 
rooted in the particle's spatial extent. 
In other words, we have $E_{\rm kin} / \Lambda \geq 1$. 
From the above results for $E_{\rm kin}$ and $(\Delta {\bf r})^2$ 
at low temperature, it is discernible that both C and Si display 
a ratio of $E_{\rm kin} / \Lambda = 1.1$. This ratio slightly exceeds 
the minimum value allowed by the uncertainty relations, and 
escalates as temperature increases, reflecting a transition toward 
more classical behavior in atomic motion.

% Table 1
\begin{table*}[t]
\caption{Elastic stiffness constants, bulk modulus, and Poisson's
ratio of $3C$-SiC, as derived from classical MD and quantum PIMD
simulations at $T$ = 0, 300 and 750~K. Data for $C_{ij}$ and $B$
are in GPa.  Statistical error bars in the last digit are given
in parenthesis.
}
\vspace{0.2cm}
\begingroup
\setlength{\tabcolsep}{10pt}
\renewcommand{\arraystretch}{1.5}
\centering
\setlength{\tabcolsep}{10pt}
\begin{tabular}{c|c c| c c| c c|}
\cline{2-7}
\multicolumn{1}{c}{} &
      \multicolumn{2}{|c|}{$T = 0$} &
      \multicolumn{2}{c|}{$T = 300$ K} &
      \multicolumn{2}{c|}{$T = 750$ K} \\ [2mm]
\cline{2-7}
     & class.  & quantum & class. & quantum & class. & quantum
  \\[2mm]
\hline
 $C_{11}$ & 452.9(5) & 428(1) & 435(1)& 425(2) & 413(2) & 409(2) \\
 $C_{12}$ & 141.1(5) & 126(1) & 129(1)& 123(2) & 116(1) & 114(2) \\
 $C_{44}$ & 246.7(5) & 236(1) & 238(1)& 235(1) & 230(2) & 228(3) \\
  $B$     & 245(1)   & 227(1) & 231(1)& 224(1) & 215(1) & 212(1)  \\
  $\nu$   &  0.238   & 0.227  &  0.230 & 0.224 & 0.219  & 0.218   \\
\hline
\end{tabular}
\endgroup
\label{el_const_simul}
\end{table*}

For an isotropic 3D harmonic oscillator characterized by a frequency 
$\omega$, the MSD for the ground state equates to 
$3 \hbar / (2 M \omega)$. Correspondingly, the kinetic energy of the 
oscillator's ground state is $(E_{\rm kin})_0 = 3 \hbar \, \omega / 4$ 
\cite{co20a}. This yields a convergence of $E_{\rm kin} / \Lambda$ 
to unity as $T \to 0$, reaching the minimum attainable value.
In the context of atomic motion in solids, the dispersion of frequencies 
can be approximated using an isotropic 3D Debye model (as referenced 
in \cite{ki96,as76}), characterized by a vibrational density of states 
$\rho(\omega) \propto \omega^2$ and a high-frequency cutoff $\omega_D$. 
In this case, for $T \to 0$ and assuming harmonic vibrations, the ratio 
$E_{\rm kin} / \Lambda$ converges to $1.125$ (which remains independent 
of $\omega_D$) \cite{he20b}, near the findings for C and Si 
atoms from our quantum simulations.

\section{Elastic constants}

In this section, we delve into the influence of nuclear quantum effects 
on the elastic stiffness constants of $3C$-SiC. These effects, akin to 
other physical observables, become pronounced for various elastic 
constants, manifesting at temperatures lower than the 
Debye temperature, $\Theta_D$.
The elastic compliance constants, $S_{ij}$, are evaluated in 
this study at various temperatures by focusing on specific components of 
the stress tensor ${ \tau_{ij} }$ during isothermal-isobaric simulations. 
For instance, when $\tau_{xx} \neq 0$ and $\tau_{ij} = 0$ for other 
components, the relationships $S_{11} = e_{xx} / \tau_{xx}$ and 
$S_{12} = e_{yy} / \tau_{xx}$ hold true. Here, $e_{ij}$ are the 
components of the strain tensor determined in the simulations 
\cite{as76,ki05,yu96}.
Similarly, the evaluation of $S_{44}$ entails the application of a shear 
stress $\tau_{xy}$, and $S_{44} = e_{xy} / \tau_{xy}$.
From these compliance constants, we derive the stiffness constants $C_{11}$ 
and $C_{12}$ using the cubic crystal relations \cite{as76,ki05}: 
$C_{11} = (S_{11} + S_{12}) / Z$ and $C_{12} = - S_{12} / Z$, where 
$Z = (S_{11} - S_{12}) (S_{11} + 2 S_{12})$. 
Additionally, $C_{44} = 1 / S_{44}$.
We note that in the context of elasticity, a hydrostatic pressure $P$ 
corresponds to $\tau_{xx} = \tau_{yy} = \tau_{zz} = -P$. 

In Fig.~8, we present the elastic stiffness constants obtained from 
our simulations of cubic SiC. The graph portrays the temperature 
dependence of $C_{11}$, $C_{12}$, and $C_{44}$ from top to bottom.
The results found from classical MD simulations are depicted using 
solid circles, designated as ``cl'', while those from 
PIMD simulations are indicated by squares and shown as ``q''.
The elastic constants derived from classical 
simulations exhibit a consistent decrease with increasing temperature 
throughout the range illustrated in Fig.~8. Notably, the extrapolation 
of these findings to $T = 0$ agrees with the data for $C_{ij}$ derived 
from the slopes of acoustic phonon bands, as discussed in Sec.~III. 
These $T = 0$ values are represented by solid diamonds on the left 
vertical axis of Fig.~8.

In Table~I, we present a compilation of stiffness constants derived
from classical and PIMD simulations at temperatures $T =$~300 and 750~K. 
Alongside these values, we provide the corresponding classical limits 
for $T = 0$, computed using the phonon bands methodology 
(Sec.~III).  Quantum values for the limit $T \to 0$ are determined 
by extrapolating finite-temperature PIMD results. A comparison between 
classical and quantum data reveals that the introduction of zero-point 
motion yields reductions of approximately 5\%, 10\%, and 4\% for 
$C_{11}$, $C_{12}$, and $C_{44}$, respectively.

Brito {\it et al.} \cite{br20} conducted an investigation into nuclear 
quantum effects in diamond utilizing path-integral Monte Carlo 
simulations with the Tersoff potential. In their analysis, these 
authors observed a reduction in the elastic constants $C_{11}$ and 
$C_{44}$ due to quantum nuclear motion, akin to the findings for
$3C$-SiC. However, their observations regarding $C_{12}$ 
demonstrated results that were nearly invariant with temperature, 
coupled with a discernible increase attributed to quantum 
motion at low $T$ (approximately 5\%). 
This stands in stark contrast to the results depicted in Fig.~8(b).
At present, the underlying cause of this disparity remains uncertain. 
It may stem from the differing characteristics of the cubic materials 
studied (diatomic versus monoatomic), or to the interatomic potential
employed in the simulations.

It is also interesting to compare our results for the elastic constants 
of cubic SiC with those obtained earlier for graphite using a similar 
procedure \cite{he21b}. For graphite, a layered material, a small 
change was found in $C_{11}$ due to quantum nuclear motion (about 1\%), 
versus 5\% obtained here for $3C$-SiC, which turns out to be similar 
to that resulting for diamond \cite{br20}.
On the contrary, $C_{12}$ undergoes a decrease in cubic SiC (10\%), 
clearly smaller than in graphite (approximately 20\%). 
These differences between cubic and layered materials shed light 
on anharmonicities associated with lattice vibrations in these 
types of solids, which manifest themselves in the effect of 
nuclear motion on magnitudes such as the elastic constants.

It is worth highlighting the correlation between the temperature-dependent 
behavior of the elastic constants and the atomic MSD.
In the case of our classical results for the three stiffness constants, 
there is a noticeable nearly linear decrease as temperature rises. 
This behavior is linked to the classical thermal motion of atoms, which 
in turn leads to a linear growth in the MSD $(\Delta {\bf r})^2$ as 
temperature is raised (as depicted in Fig.~7).
Conversely, when examining data obtained from PIMD simulations, 
a pronounced reduction is observed in the low-temperature elastic constants. 
This reduction is attributed to the effects of zero-point delocalization, 
evident in the finite MSD at $T = 0$.
As $T$ is elevated, a convergence between classical and quantum data 
for $C_{ij}$ takes place, much like the convergence observed in 
the MSD. A similar 
pattern is also observed in the classical and quantum data for the 
bulk modulus, as illustrated in Sec.~VIII.

% Table 2
%~\vspace{-8cm}
\begin{table}
\caption{Elastic stiffness constants, bulk modulus $B$, and
Poisson's ratio $\nu$ of $3C$-SiC derived from results of
experimental techniques by different authors at ambient
conditions. In each case, the bulk modulus is obtained
from the elastic constants by means of Eq.~(\ref{bulkm}).
$C_{ij}$ and $B$ are expressed in GPa.
Error bars in the last digit, when available,  are given
in parenthesis.
}
\vspace{0.2cm}
\centering
\setlength{\tabcolsep}{10pt}
\begin{tabular}{c c c c}
     & Zhuravlev \cite{sc-zh13} &  Lee \cite{sc-le82} &
       Lambrecht \cite{sc-fe68,sc-la91}
  \\[2mm]
\hline  \\[-2mm]
  $C_{11}$  &  391(11) &  363    &  390    \\[2mm]
  $C_{12}$  &  122(6)  &  154    &  142    \\[2mm]
  $C_{44}$  &  253(6)  &  149    &  256    \\[2mm]
    $B$     &  218(1)  &  224    &  225    \\[2mm]
    $\nu$   &  0.238   &  0.298  &  0.267  \\[2mm]
\hline  \\[-2mm]
\end{tabular}
\label{el_const_exp}
\end{table}

In Table~II, we give a compilation of $C_{ij}$ values extracted 
from experimental data as reported by various authors 
\cite{sc-zh13,sc-le82,sc-fe68,sc-la91}. A certain degree of 
variation is evident in the experimental results obtained under ambient 
conditions, especially for $C_{44}$. In particular, the value 
given in Ref.~\cite{sc-le82} is notably lower than the other reported 
values for this stiffness constant. 
Moving on to Table~III, we present a comprehensive overview of elastic 
constants for cubic silicon carbide calculated by using different
computational methodologies.
Some calculations were conducted within the framework of DFT, 
incorporating both LDA \cite{sc-wa96} and generalized-gradient 
approximation (GGA) \cite{sc-le15b,sc-ra21}. 
Other results were obtained through lattice-dynamics calculations 
\cite{sc-le82}, the utilization of effective interatomic potentials 
\cite{sc-va15}, and MD simulations \cite{sc-sh00}.

% Table 3
\begin{table*}[t]
\caption{Elastic stiffness constants, bulk modulus, and Poisson's
ratio of $3C$-SiC found from several calculations based on
density-functional theory with LDA and GGA (PBEsol and PBE), as well
as lattice dynamics (LD), an effective potential (EP), and MD
simulations.  Data for $C_{ij}$ and $B$ are given in GPa.
The bulk modulus is calculated in each case from
the elastic constants using Eq.~(\ref{bulkm}).
}
\vspace{0.2cm}
\centering
\setlength{\tabcolsep}{10pt}
\begin{tabular}{c c c c c c c}
 & LD \cite{sc-le82} & EP \cite{sc-va15} & MD \cite{sc-sh00} &
   PBEsol \cite{sc-le15b} & PBE \cite{sc-ra21} & LDA \cite{sc-wa96}
  \\[2mm]
\hline  \\[-2mm]
$C_{11}$  &  371   & 371.1  &  390  & 390.0  & 382.9 & 384   \\[2mm]
$C_{12}$  &  169  &  223.4  &  144  & 137.6  & 126.9 & 132   \\[2mm]
$C_{44}$  &  176  &  279.3  &  179  & 246.3  & 240.9 & 241   \\[2mm]
  $B$     &  225  &  273    &  225  & 224.2  & 212.2 & 216   \\[2mm]
  $\nu$   & 0.313 &  0.376  & 0.270 & 0.261  & 0.249 & 0.256 \\[2mm]
\hline  \\[-2mm]
\end{tabular}
\label{el_const_calc}
\end{table*}

The Poisson's ratio, $\nu$, is a parameter that characterizes the 
relationship between transverse and longitudinal strains under an 
applied stress. For cubic SiC, we calculate it
as $\nu = - S_{12} / S_{11}$ \cite{sc-la91}.
In Table~I, we furnish values of Poisson's ratio derived from the 
elastic constants obtained through our classical and quantum 
simulations. It is noteworthy that as temperature increases, 
$\nu$ shows a reduction.
For $T \to 0$, the influence of zero-point motion causes a decline 
in $\nu$ from 0.238 to 0.227, signifying a reduction of 5\%. 
At $T$ = 300 K, the classical and quantum values of $\nu$ stand at 
0.230 and 0.224, respectively, indicating a decrease of about 3\% 
attributed to quantum motion.

Table~II provides the Poisson's ratio values acquired from experimental 
investigations. Notably, the value derived from the data reported 
by Zhuravlev {\it et al.} 
\cite{sc-zh13} closely approximates our results, whereas the other two 
values are relatively higher \cite{sc-le82,sc-la91}.
Table~III presents the Poisson's ratio values attained through various 
theoretical methods. While some dispersion exists in these outcomes, 
the lowest values correspond to DFT calculations, both within LDA and 
GGA. These DFT-based results are somewhat higher than our classical 
result at $T = 0$ ($\nu = 0.238$).

\begin{figure}
\vspace{-7mm}
\includegraphics[width=7cm]{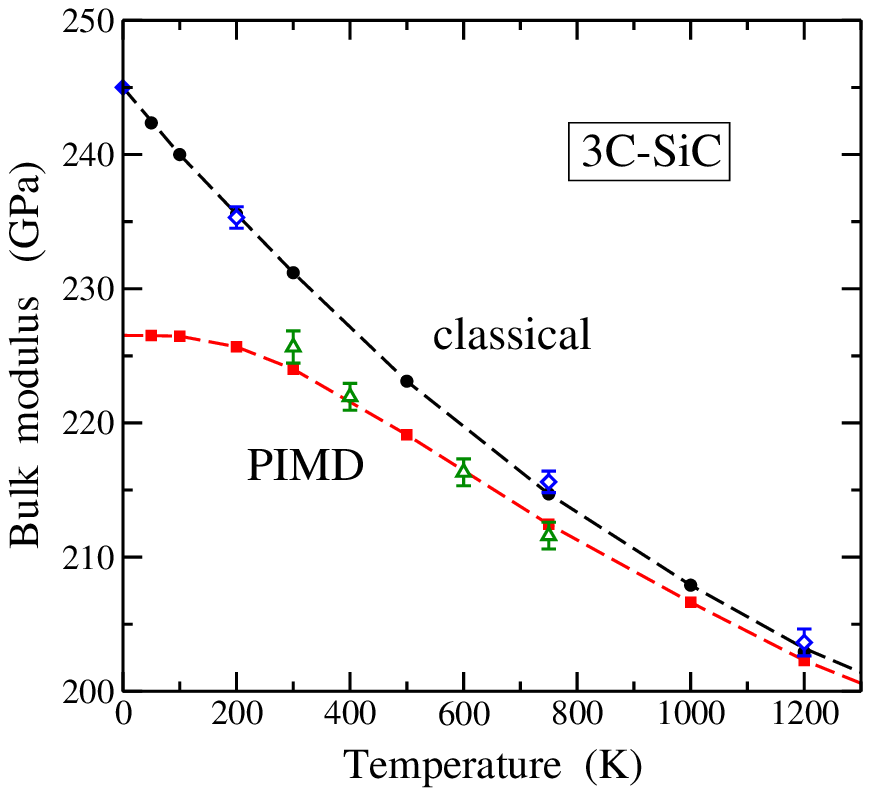}
\vspace{-5mm}
\caption{Temperature dependence of the bulk modulus of $3C$-SiC,
obtained from the elastic constants derived from classical MD
(solid circles) and PIMD simulations (solid squares).
Error bars of these data are in the order of the symbol size.
A solid diamond indicates the classical value for $T = 0$.
Open symbols represent data for the bulk modulus found using
the fluctuation formula in Eq.~(\ref{b_fluc}): diamonds, classical
simulations; triangles, PIMD simulations.
}
\label{f9}
\end{figure}

\section{Bulk modulus}

In this section, we focus on the calculation of the isothermal bulk
modulus, $B = - V (\partial P / \partial V)_T$, within our computational
approach.  As depicted in Fig.~9, we showcase the temperature-dependent 
behavior of $B$ computed through the elastic constants, using 
Eq. (\ref{bulkm}).  Solid circles and squares denote the outcomes 
obtained from classical and PIMD simulations, respectively. 
The diamond marker positioned at $T = 0$ 
signifies the classical value, derived from the phonon bands using the 
harmonic approximation, as detailed in Sec.~III.

The classical results exhibit a nearly linear decrease 
at low temperatures, accompanied by a substantial slope 
$\partial B / \partial T$, which becomes less negative as temperature 
escalates. Notably, at temperatures of 300 and 1200~K, there is a 
respective reduction in $B$ by 6\% and 17\% when compared to 
the low-temperature limit.
For decreasing temperature, the extrapolation to $T = 0$ yields values of 
$B$ = 245(1) GPa and 227(1) GPa for classical and quantum simulations, 
respectively. This signifies a 7\% decline in the bulk modulus due to 
the influence of atomic zero-point motion.
The two sets of results converge as temperature increases, with a 
difference between them of roughly 0.6\% at 1000 K.  A salient aspect 
is that the quantum outcomes fulfill the requirement of the third law 
of Thermodynamics \cite{ca85,be00b}, namely that 
$\partial B / \partial T = 0$ as temperature approaches zero. 
Conversely, this law is not upheld by the classical data, where 
$\partial B / \partial T < 0$ remains evident down to $T = 0$.

Table~I provides a comprehensive overview of the bulk modulus for 
$3C$-SiC. These values are derived by means of Eq.(\ref{bulkm}), using 
the stiffness constants obtained through classical and quantum 
simulations at temperatures of 300 and 750~K, as well as the 
zero-temperature limit. Additionally, Tables~II and III furnish 
bulk modulus values acquired through experimental and theoretical 
procedures, respectively.
The experimental data for the bulk modulus of cubic SiC obtained by 
various authors under room temperature conditions closely align with 
our PIMD simulation result of $B = 224$~GPa at 300~K. Our {\em ab-initio} 
DFT calculations, carried out without considering nuclear quantum 
effects at $T = 0$, yield a value of $B = 224$~GPa, consistent with the 
findings of Lee and Yao \cite{sc-le15b}, who employed the PBEsol 
generalized-gradient approximation. Other DFT calculations report 
values of 212 GPa \cite{sc-ra21} and 216 GPa \cite{sc-wa96}.

In the context of our classical and quantum atomistic simulations, 
additional insight into the behavior of the bulk modulus at low 
temperature can be garnered by means of the fluctuation 
formula \cite{la80}:
\begin{equation}
  B = \frac{k_B T V_c}{(\Delta V_c)^2} =
        \frac{k_B T V}{N (\Delta V)^2}  \; ,
\label{b_fluc}
\end{equation}
where $V_c = N V$ signifies the volume of the simulation cell
and $\Delta V$ represents the volume fluctuations.
For a given $N$, it is observed that $(\Delta V)^2 \sim T$ at low 
temperatures, regardless of the classical or quantum nature of
the simulations.  Consequently, the dependence of $B$ on 
temperature is predominantly governed by the behavior of the 
function $V(T)$.
In a classical model, the volume $V$ increases linearly as temperature 
rises: $(V - V_0) \sim T$. Conversely, the quantum behavior shows
$\partial V / \partial T = 0$ in the low-temperature limit. 
This distinction is consistent with the observed tendencies at 
low $T$: $B$ decreases linearly for rising $T$ in classical 
simulations, and $\partial B / \partial T \to 0$ for 
quantum simulations.

Upon analyzing Eq.~(\ref{b_fluc}) alongside the actual $B$ values 
obtained from our simulations, we observe that the change in $B$ 
attributed to nuclear quantum motion at low temperatures primarily 
corresponds to an increase in $\Delta V$ compared to classical outcomes. 
Remarkably, our PIMD simulations yield an approximately 7\% increase 
in $(\Delta V)^2$ compared to its classical counterpart, mirroring 
the shift observed in the bulk modulus $B$. The variation in the 
mean volume, $V$, remains below 1\% (refer to Sec.~V).

Eq.~(\ref{b_fluc}) introduces an alternative approach for computing 
the bulk modulus through simulations. We have verified that this method 
produces results that align, within error bars, with those obtained 
from elastic constants employing Eq.(\ref{bulkm}). Several data points 
obtained using the fluctuation formula are illustrated in Fig.~9, 
denoted by open symbols: diamonds represent classical outcomes, 
while triangles depict quantum data. It is important to note that 
the error bars associated with this method are comparatively larger 
than those arising from the determination through elastic constants.

\begin{figure}
\vspace{-7mm}
\includegraphics[width=7cm]{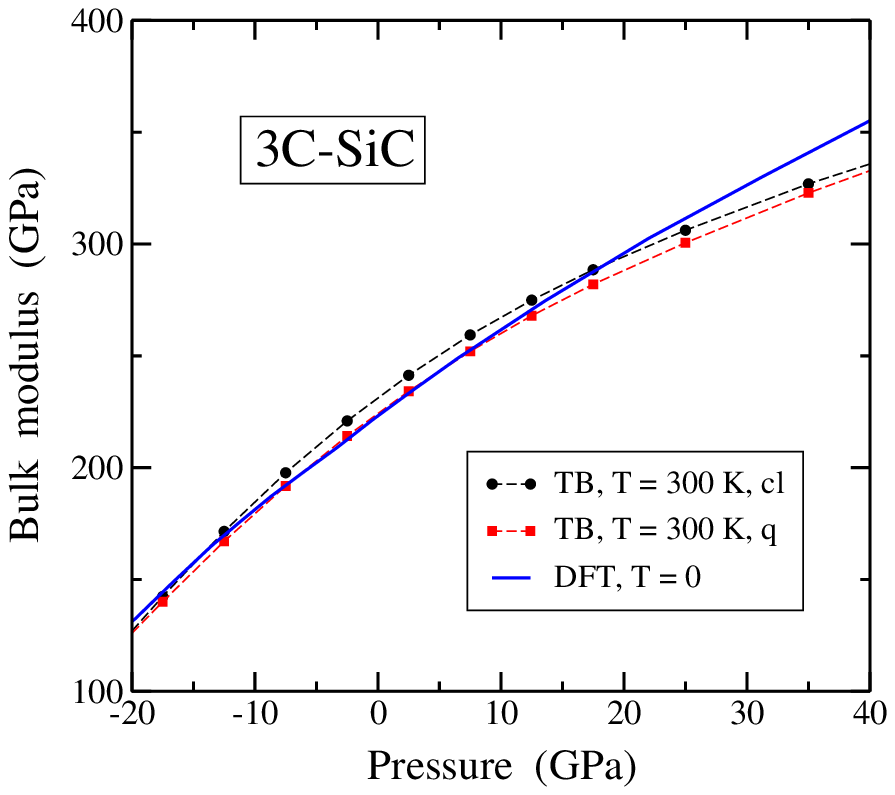}
\vspace{-5mm}
\caption{Bulk modulus of $3C$-SiC, as derived from classical
MD (circles) and PIMD simulations (squares) at $T$ = 300~K
for several hydrostatic pressures.
Dashed lines are guides to the eye.
The solid line represents the results of DFT calculations
at $T = 0$.  Labels ``cl'' and ``q'' refer to classical and
quantum data, respectively.
}
\label{f10}
\end{figure}

In Fig.~10, we have depicted the relationship between the bulk 
modulus and hydrostatic pressure. The symbols correspond to 
results from classical (circles) and PIMD simulations (squares) 
at $T =$~300~K. The solid line represents the results obtained 
from DFT calculations at $T = 0$. The TB results at $T = 0$ slightly 
surpass the classical data at $T = 300$~K, and are omitted for 
the sake of clarity. As the compressive pressure reaches large 
values, the DFT result becomes larger that derived from the 
TB model. This disparity between the two datasets escalates with 
increasing compressive pressure. Specifically, at $P = 40$~GPa, 
the difference is approximately 6\%.

The DFT results exhibit a pressure derivative of the bulk modulus, 
$B' = \partial B / \partial P$, which amounts to 3.8 at $P = 0$. 
In the case of TB data, we observe values of 4.1 and 4.0 for 
classical and quantum simulations at $T = 300$~K, respectively. 
Previous DFT calculations have consistently yielded $B'$ values 
ranging between 3.7 and 3.9 \cite{sc-le15b,sc-ra21,sc-wa96,sc-ka94}. 
Moreover, various experimental studies have reported values of 
the pressure derivative at room temperature spanning from 
3.6 to 4 \cite{sc-zh13,sc-st87,sc-yo93}.

It is worth noting that the difference between classical and 
quantum results for the bulk modulus, $\delta B$, diminishes as 
the pressure increases. 
In terms of Eq.~(\ref{b_fluc}), this is due to a reduction
in the difference between quantum and classical MSD
$(\Delta V)^2$ for rising hydrostatic pressure, which is
associated to an increase in the mean frequency $\overline{\omega}$.

\section{Summary}

PIMD simulations provide a robust framework to quantitatively assess 
nuclear quantum effects on the structural and elastic properties of 
solids beyond harmonic or quasi-harmonic approximations.
In the case of silicon carbide, quantum corrections manifest 
significantly, especially at temperatures below 400~K.

The employment of an effective tight-binding  Hamiltonian has
enabled precise exploration of the interplay between nuclear
quantum motion and anharmonicity, factors that appreciably influence
the material's behavior at low temperature. 
The quality of the TB model in describing the physical properties 
of $3C$-SiC has been established through a direct comparison with
DFT calculations at $T = 0$. The pressure-volume equations of state
derived from both methods are close 
to one another in a wide range of pressures, including tensile
and compressive stress. For $P \gtrsim 30$~GPa, the bulk modulus
derived from TB calculations is smaller than that found using
the DFT procedure.

The comparison between PIMD and classical MD simulations emphasizes 
the importance of nuclear quantum effects in understanding the behavior 
of SiC. Furthermore, our exploration of anharmonicity using
a harmonic approximation helps to deepen our comprehension of the
material's vibrational properties.

The process of quantizing lattice vibrations engenders discernible
alterations in the volume and elastic properties of $3C$-SiC as 
compared to a classical model.  
The quantum zero-point expansion of the volume amounts to about 1\% 
respect the classical prediction.
In addition, low-temperature quantum corrections in the elastic 
stiffness constants $C_{11}$, $C_{12}$, and $C_{44}$, amount to
reductions of 5\%, 10\%, and 4\%, respectively.
Similar decreases are found for the bulk modulus (7\%) and
Poisson's ratio (5\%).
Collectively, our findings underscore the overarching trend: $3C$-SiC 
exhibits a ``softer'' character than predicted by classical simulations,
in particular at low temperature.

We highlight the congruence between our simulation outcomes 
and the principles of thermodynamics, in particular the third law, 
which applies for $T \to 0$. Specifically, this alignment implies 
in our case that the temperature derivatives of the mechanical 
properties (i.e., elastic constants and bulk modulus) must tend 
toward null values in the low-temperature limit, contrary to 
the results of classical simulations.

In contrast to preceding works in this field, our study extends 
the pressure range to encompass tensile stress ($P < 0$), thereby 
enabling an examination of SiC within a metastable region of its 
phase diagram. Additionally, our investigation benefits from a robust 
foundation for the PIMD simulations, established from direct comparison 
between TB outcomes and DFT calculations at $T = 0$. 
Consideration of silicon carbide facilitates an evaluation of 
the collective impact of anharmonicity and quantum nuclear motion on 
silicon and carbon atoms within a binary compound, surpassing previous 
examinations of related monoatomic materials. This is particularly 
noteworthy in our analysis of MSDs and kinetic energy pertaining 
to Si and C atoms in $3C$-SiC.

In summary, our study emphasizes the advantages of
including a quantum description of atomic nuclei into silicon
carbide research, significantly impacting our comprehension of
its thermal and mechanical attributes.
As we broaden our exploration of nuclear quantum effects, it
opens the door to an understanding of the involved
interralation between phonon quantization and anharmonicity
in the physical properties of diverse materials.

\begin{acknowledgments}
This work was supported by Ministerio de Ciencia e Innovaci\'on
(Spain) through Grants PGC2018-096955-B-C44 and PID2022-139776NB-C66.
\end{acknowledgments}

%  -----------------------------------------------------------------

\appendix

\section{Perturbed harmonic oscillator}

   A qualitative understanding of the low-temperature decrease in
kinetic energy with respect to the value expected in a harmonic
approximation can be obtained by analyzing changes
in kinetic and potential energy by standard time-independent
perturbation methods.
We consider a one-dimensional harmonic oscillator with potential
energy $V(x) = \frac12 m \omega^2 x^2$, and a perturbation
\begin{equation}
     W(x) = A x^3 + B x^4  \; .
\end{equation}
In static perturbation theory, the perturbed ground state is given
to first order by \cite{la65,co20b}:
%\begin{equation}
\begin{eqnarray}
 E_0^1 \; & =  & \;  \langle \psi_0^0| V(x) + W(x) |\psi_0^0 \rangle  \nonumber  \\
	&  = & \; \frac{1}{2} \hbar \omega +
    \frac {3}{4} \left( \frac {\hbar}{m \omega} \right)^2 B +
    o(A^2,B^2)   \;  .
\label{e01}
\end{eqnarray}
%\end{equation}
The superscript in the wave function $|\psi_0^0 \rangle$
indicates the unperturbed state.  Note that, to first-order, 
the contribution of the cubic term $A x^3$ vanishes.

By calculating separately changes in the kinetic and potential energy
in the perturbed oscillator, one finds that the change in
ground-state energy, $(\delta E)_0^1 = E_0^1 - E_0^0$, 
is due to a variation of the kinetic
energy, and the potential energy keeps constant for first-order
perturbation.
This can be seen by calculating the expectation value of the corresponding
operators in the  first-order-corrected ground state \cite{la65,co20b}:
%\begin{equation}
\begin{eqnarray}
 |\psi_0^1 \rangle  \; & = & \; |\psi_0^0 \rangle
   - \; \frac{3}{2 \sqrt{2}} C_1 A |\psi_1^0 \rangle
   - \; \frac{3}{2 \sqrt{2}} C_2 B |\psi_2^0 \rangle    \nonumber  \\
  & - & \; \frac{1}{2 \sqrt{3}} C_1 A |\psi_3^0 \rangle
   - \; \frac{\sqrt{3}}{4 \sqrt{2}} C_2 B |\psi_4^0 \rangle  \;  ,
\label{psi01}
\end{eqnarray}
%\end{equation}
where $|\psi_n^0 \rangle$ refers to the $n$'th unperturbed wave
function, $C_1 = (\hbar / m^3 \omega^5)^{1/2}$ and
$C_2 = \hbar / m^2 \omega^3$.

Then, for the potential and kinetic energy in the perturbed ground
state, $|\psi_0^1 \rangle$, we have:
\begin{equation}
  \langle E_{\rm pot} \rangle_0^1 \; = \;
        \langle \psi_0^1| V(x) +  W(x) |\psi_0^1 \rangle
       \;  =  \; \frac{1}{4} \hbar \omega + o(A^2,B^2) \;  ,
\label{ep0}
\end{equation}
and
\begin{eqnarray}
 \langle E_{\rm kin} \rangle_0^1 \; & = & \;
 \langle \psi_0^1| \frac {p^2}{2 m} |\psi_0^1 \rangle \; = \nonumber \\
	\; & = & \;  \frac{1}{4} \hbar \omega +
        \frac {3}{4} \left( \frac {\hbar}{m \omega} \right)^2 B
          + o(A^2,B^2) \;  .
\label{ek0}
\end{eqnarray}

%  -----------------------------------------------------------------

\section{Zero-temperature volume}

  For a given volume, the energy $E$ at $T = 0$ may be written
as $E = E_{\rm cl} + E_{\rm ZP}$, where $E_{\rm cl}$ is the
classical energy for motionless atoms (given in Fig.~3) 
and $E_{\rm ZP}$ is the zero-point energy per atom:
\begin{equation}
   E_{\rm ZP} = \frac12  \int\limits_0^{\omega_{\rm max}}
            \frac12 \hbar  \omega \, g(\omega) \, d\omega \;
          =  \frac32 \hbar \, \overline{\omega}  \; , 
\label{ezp}
\end{equation}
where $\overline{\omega}$ is the mean frequency:
\begin{equation}
  \overline{\omega} = \frac16 \int\limits_0^{\omega_{\rm max}} 
	    \omega \, g(\omega) \, d\omega \; ,
\label{ommed}
\end{equation}
and the factor $1/6$ in front of the integral comes from 
the normalization condition in Eq.~(\ref{intg}).
The Gr\"uneisen parameter $\overline{\gamma}$ is defined as
\begin{equation}
 \overline{\gamma} = - \frac {\partial ({\rm log} \, \overline{\omega}) }  
          {\partial ({\rm log} \, V) }  =
     - \frac{V}{\overline{\omega}} \frac {\partial \overline{\omega} }
	     {\partial V }  \; ,
\label{gamma2}
\end{equation}
and using Eqs.~(\ref{ezp}) and (\ref{gamma2}), we have
\begin{equation}
     \frac {\partial E_{\rm ZP}} {\partial V} = \frac{3 \hbar}{2} 
	 \frac {\partial \overline{\omega} } {\partial V } =
	 - \frac {\overline{\gamma} E_{\rm ZP}} {V}  \; .
\label{dezp2}
\end{equation}
For equilibrium at $P = 0$, the volume derivative of the energy
$E$ has to vanish, i.e.:
\begin{equation}
   \left( \frac{\partial E_{\rm cl}} {\partial V} \right)_{V_{min}} =
 - \left( \frac{\partial E_{\rm ZP}} {\partial V} \right)_{V_{min}}
    \approx  \frac{\overline{\gamma} E_{\rm ZP}} {V_0}  \; ,
\label{dev2}
\end{equation}
which gives the volume $V_{min}$ corresponding to the quantum
ground state.

We finally evaluate the zero-point volume expansion due to quantum
motion, i.e., the difference $V_{\rm min} - V_0$.
Taking into account that 
\begin{equation}
  E_{\rm cl} = E_0 + \frac12 \frac{B_0}{V_0} (V - V_0)^2 + ... \, ,
\end{equation}
a good estimation of the volume increase is given by the expression
\begin{equation}
   \left( \frac{\partial E_{\rm cl}}
         {\partial V} \right)_{V_{\rm min}}  =
         \frac{B_0}{V_0} (V_{\rm min}- V_0) \; .
\label{dev3}
\end{equation}
As a result of Eqs. (\ref{dev2}) and (\ref{dev3}), we have:
\begin{equation}
 V_{\rm min}  = V_0 + \frac{ \overline{\gamma} E_{\rm ZP} } 
	 {B_0} \; .
\label{vmin2}
\end{equation}

%  -------------------------------------------------------------------

%    BIBLIOGRAPHY

%

\end{document}